# Ultra-high mobility semiconducting epitaxial graphene on silicon carbide

Jian Zhao[1#], Peixun Ji[1#], Yaqi Li[1#], Rui Li[1#], Kaiming Zhang[1], Hao Tian[1], Kaichen Yu[1], Boyue Bian[1], Luzhen Hao[1], Xue Xiao[1], Will Griffin[2], Noel Dudeck[2], Ramiro Moro[1],
Lei Ma[1§*], Walt A. de Heer[1,2§*]

1. Tianjin International Center for Nanoparticles and Nanosystems. Tianjin University. Tianjin. PR. China
2. School of Physics, Georgia Institute of Technology, Atlanta, Georgia, United States

Corresponding Authors:
*Lei.ma@tju.edu.cn.
*walt.deheer@gatech.edu
#,§These authors contributed equally

## Abstract

Graphene nanoelectronics' potential was limited by the lack of an intrinsic bandgap[1] and attempts to tailor a bandgap either by quantum confinement or by chemical functionalization failed to produce a semiconductor with a large enough band gap and a sufficient mobility. It is well known that by evaporating silicon from commercial electronics grade silicon carbide crystals an epitaxial graphene layer forms on the surfaces [2]. The first epigraphene layer to form on the silicon terminated face, known as the buffer layer, is insulating. It is chemically bonded to the SiC and spectroscopic measurements [3] have identified semiconducting signatures on the microscopic domains. However, the bonding to the SiC is disordered and the mobilities are small. Here we demonstrate a quasi-equilibrium annealing method that produces macroscopic atomically flat terraces covered with a well ordered epigraphene buffer layer that has a 0.6 eV bandgap. Room temperature mobilities exceed 5000 cm$^2$/Vs which is much larger than silicon and 20 times larger than the phonon scattering imposed limit of current 2D semiconductors. Critical for nanotechnology, its lattice is aligned with the SiC substrate, it is chemically, mechanically, and thermally robust, and it can be conventionally patterned and seamlessly connected to semimetallic epigraphene making semiconducting epigraphene ideally suited for nanoelectronics.

**Main**

The graphene revolution was originally driven by the search for new electronic materials that could succeed silicon [4]. Graphene, which is a semimetal (i.e., a gapless semiconductor) was a likely candidate [5] following predictions that, due to quantum confinement, graphene nanoribbons can be semiconductors [6] [7]. Efforts to produce high quality semiconducting ribbons were not successful [8], so research focused on altering the electronic structure of graphene chemically to produce a viable semiconductor [9]. However, these efforts also had limited success [10] and interest shifted away from graphene towards 2D materials that are intrinsically semiconducting [1, 11]. Here we show that well-annealed graphene on a specific silicon carbide crystal face is an extremely high mobility 2D semiconductor.

Epigraphene is graphene that spontaneously forms on silicon carbide crystals when silicon sublimates from the surface at high temperatures resulting in a carbon rich surface that recrystallizes into graphene[2, 12]. Epigraphene on the silicon terminated face of h-SiC -the h-SiC (0001) face- has been the focus of much research[13] [2]. The first graphene layer to grow on this surface is bonded to the SiC surface and is possibly a semiconductor[14, 15]. This buffer layer has the graphene lattice structure. It is rotated by 30° with respect to the SiC lattice to produce a quasi-periodic $SiC_{6x6}$ superlattice with a lattice constant of 1.85 nm.

Angle resolved photoelectron spectroscopy (ARPES) gave compelling evidence that the buffer layer is indeed a semiconductor [3, 14, 16] however the mobility µ of the buffer layer was found to be only µ=1 $cm^2V^{-1}s^{-1}$ [17] (Sup Mat), which is small compared with other 2D semiconductors that have room temperature mobilities up to about 300 $cm^2V^{-1}s^{-1}$ [1, 18].

X-ray reflection studies of the buffer layer show that the underlying SiC surface is significantly depleted of silicon[19, 20]. This is not too surprising because the buffer layer is produced by the thermal depletion of silicon by sublimation. When Si is completely depleted, the buffer layer becomes graphene under which a new buffer layer crystalizes[14]. Consequently, while the buffer layer has essentially a perfect graphene structure, the bonding to the substrate is disordered and the mobilities are small [21] [17].

Here we demonstrate a quasi-equilibrium production method that produces high quality semiconducting epigraphene (SEG) on macroscopic domains with a band gap of 0.6 eV and room temperature mobilities up to µ=5500 $cm^2V^{-1}s^{-1}$ which is 3 times greater than silicon and a factor of 20 greater than theoretically possible with any other 2D semiconductor investigated to date[18]. In addition, SEG is atomically registered with the SiC lattice and patternable using conventional methods, making it an ideal platform for 2D nanoelectronics [5].

**SEG production**

Conventional epigraphene and buffer layer is grown in a confinement-controlled sublimation (CCS) furnace [22] (Fig 1a-b) where a 3.5 mm x 4.5 mm semi-insulating SiC chip is annealed in a cylindrical graphite crucible in a 1 bar Ar atmosphere at temperatures ranging from 1300 C to

1600 C (Fig. 1c). The crucible is supplied with a small leak. The rate silicon escapes from the crucible determines the rate at which graphene forms on the surface. In this way, the growth temperature and the graphene formation rates are controlled.

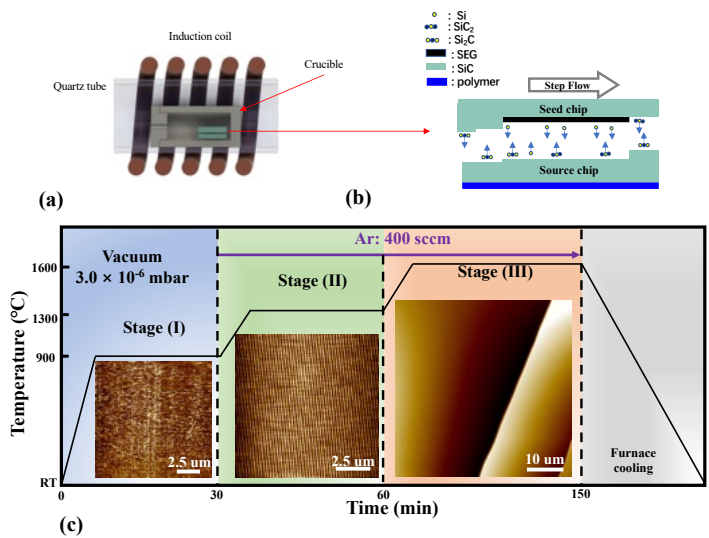

**Fig. 1** SEG production, (**a**) Schematic diagram of a CCS furnace with two 3.5 mm x 4.5 mm SiC chips inside a closed cylindrical graphite crucible that is supplied with a leak inside a quartz tube. The crucible is inductively heated by high frequency currents through a coil. (**b**) The two chips are stacked with the C-face of the bottom chip (Source) facing the Si-face of the top chip (Seed). At high temperatures a slight temperature difference between the chips causes a net mass flow from the bottom chip to the top chip resulting in the growth of large terraces on the seed chip by step flow growth on which a uniform SEG film grows. (**c**) SEG is grown in three stages. In stage (I) in vacuum the chip is heated to 900 C for about 25 min to the surface; Heating to 1300 C for about 25 min. 1 bar of Ar produces a regular array of bilayer SiC steps and ≈ 0.2 μm wide terraces. SEG coated (0001) terraces grow in Stage (III) at 1600 C in 1 bar of Ar step bunching and step flow produce large atomically flat terraces on which a buffer layer grows in quasi-equilibrium conditions established between the C-face and the Si-face. The large SEG coated (0001) terraces are explained in terms of its very large stability.

If the leak is sealed, then graphene growth is strongly suppressed. Graphene growth is further suppressed in the so-called sandwich or face-to-face method (Sup Mat), [17] [23] where two chips are stacked, typically with the Si-face of one chip facing the Si-face of the other. In 1 bar of Ar, virtually no silicon can diffuse out of the micron scale gap between the chips so that the 1:1 Si:C ratio is maintained, even at high temperatures where the Si evaporation rates from the surfaces are high. Under these conditions significant step flow and step bunching is observed[24] (Fig 1c). Step bunching is the process where substrate surface steps due to the unavoidable slight miscut of the crystal which is nominally cut along the (0001) face, merge to produce large atomically flat (0001) terraces bounded by proportionally high steps.

We observe that when the Si-face opposes a C-face then large atomically flat terraces that are uniformly covered with a buffer layer grow at temperatures around 1600 C in a 1 bar ultrapure Ar atmosphere (Fig. 1c). While the Si vapor pressure dominates, at T>1600 C the vapor pressures of $Si_2C$ and $SiC_2$ are already sufficient to promote significant SiC transport from the C-face to the Si-face[25]. This process contrasts the conventional nonequilibrium CCS method where the Si-face is continuously depleted of Si. The original experiments were conducted by the Tianjin group using semi-insulating SiC chips where the bottom chip (see Fig. 1) was coated with a polymer to produce large SEG coated (0001) terraces. The graphitized polymer likely causes the bottom chip to become slightly hotter (see below). Samples produced by this method were used in the transport measurements reported here.

The production method is similar to the physical vapor sublimation process for silicon carbide crystal production where source SiC crystallites at high temperatures in an Ar filled graphite crucible sublime and the vapors condense on a cooler SiC seed crystal to produce perfect SiC crystals with large terraces [26], so the large terrace formation can be expected in our system as well.

The most relevant parameters are the temperature T, the temperature difference between the chips $\Delta T$, and the annealing time t, which for T=1600-1700 C is typically t=1-2 hours. The temperature difference $\Delta T$ depends on the crucible design, estimated to be on the order of 10 C, to provide a vapor pressure differential between the 2 chips required for sufficient mass transport [see Methods]. Guided by these principles, alternative crucible designs, chip configurations and annealing processes were tested that do not require a polymer coated chip. (see also Sup. Mat, and Methods).

Summarizing, we found evidence that a thin Si film forms on the hotter C-face, in the C-face to Si-face configuration while large SEG coated (0001) facets grow on the Si-face. Therefore, Si which is missing from the Si face, may in fact condense on the C-face to restore the overall stoichiometry. In any case, only SEG is formed and there is no evidence for graphene. We also find that large SEG coated (0001) terraces also form on the Si-face, when the temperature gradient is inverted so that the Si-face is hotter than the C-face and mass transport is from Si-face source to the C-face seed. Apparently in this inverted crystal growth, the substrate steps evaporate from the source to leave large (0001) terraces on the Si-face. Furthermore, in a Si-face to Si-face configuration, we find that the large SEG coated terraces form on the hotter Si face and not on the cooler Si face. Moreover, in experiments using a single chip and where bulk silicon is introduced into the crucible so produce a silicon vapor saturated environment in the crucible, the Si face is of the chip is partly coated with SEG (see Methods) and no graphene is found on the C face.

From these experiments we conclude that the SEG coated (0001) facet is extremely stable, more stable than any other SiC facet and specifically, more stable than a bare (0001) face, implying that in principle it should be possible to produce wafer scale single crystal SEG.

**SEG characterization**

SEG is investigated on all relevant length scales. On the 100 nm to the 1 mm scale, scanning electron microscopy (SEM) can provide a high contrast that distinguishes bare SiC, SEG and graphene [27] (Fig. 2a). On the nanometer scale, graphene and SEG are also readily identified in scanning tunneling microscopy (STM) by its SiC$_{6x6}$ modulation (Fig. 2b). Low energy electron diffraction (LEED) is used to identify SEG and to verify its atomic registration with the SiC substrate[2] (Fig 2c). LEED is also used to distinguish SEG from graphene [2]. Raman spectroscopy (1 μm to 100 μm) is very sensitive to graphene and SEG, and traces of graphene are easily identified by its intense characteristic 2D peak[28, 29] (Fig. 2c). Lateral force microscopy (LFM) distinguishes SEG from SiC and graphene in micron scale scans. Atomic force microscopy (AFM), SEM and optical microscopy can identify surface steps. AFM is used to measure the amplitude of the steps (Fig. 6). Using a combination of these probes we find that in a wide range of production temperatures graphene is absent to any detectable level, even on substrate steps where it is readily formed in the conventional CCS method. Figure 2e shows cryogenic scanning tunneling spectroscopy which maps the density of states (DoS) of SEG as a function of the Fermi energy. The image shows a well-defined band gap of 0.6 eV. There are no detectable states in the band gap in contrast to buffer layer samples produced by conventional sublimation methods [21].

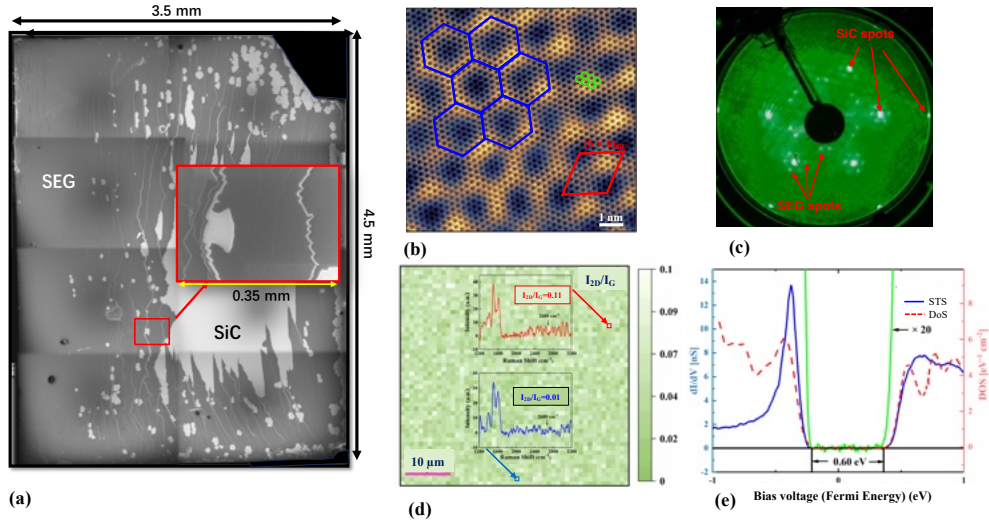

**Fig. 2**
SEG characterization demonstrating high coverage of well-ordered, graphene free, crystallographic aligned SEG, with a well-defined bandgap. (**a**) Composite electron microscope image of a full 3.5 mm x 4.5 mm wafer. The SEM is tuned to provide a vivid contrast between SiC (white areas) and SEG (grey areas). Approximately 80% of the surface is covered with SEG. Graphene would show up as very dark patches (the black spots seen here are dust particles). The largest step-free areas are about 0.5 mm by 0.3 mm. (**b**) Low temperature, atomic resolution STM image of SEG showing the graphene honeycomb lattice (green) that is spatially modulated with a $(6 \times 6)_{SiC}$ super-periodic structure (red rhombus; purple hexagons) that corresponds to the SEG height modulation due to the covalent bonding to the substrate. (**c**) LEED of SEG showing the characteristic $6\sqrt{3} \times 6\sqrt{3} R30°$ diffraction pattern of the SEG lattice, which reveals its graphene-crystal structure and the crystallographic alignment of the SEG with respect with the SiC substrate atoms. There is no trace of graphene which is invariably abundant in conventionally produced buffer layer samples. (**d**) Raman map of a 50 μm by 50 μm area with a 1 μm resolution measuring the intensity ratio $I_{2D}/I_G$ of the intensity at 2680 cm$^{-1}$ and at 1620 cm$^{-1}$. For graphene $I_{2D}/I_G \approx 2$. The red arrow corresponds to the red spectrum taken at the spot where the intensity ratio is the largest in the 2500 spectra in the map demonstrating the absence of any graphene on the surface as confirmed with other probes. (**e**) Low temperature STS of SEG, showing the 0.6 eV band gap of SEG (blue line) compared with the calculated DoS of SEG (red dashed line). There is no measurable intensity in the gap indicating a low density of impurity states.

## SEG transport properties

A series of transport measurements on SEG Hall bars that were patterned using two different methods. The samples were p-doped in ambient air in pure oxygen or in pure $O_2$, and with ultraviolet (UV) radiation. In this way room temperature charge densities n from n=4x10$^{12}$ cm$^{-2}$ to

$4 \times 10^{13}$ cm$^{-2}$ were achieved. We are specifically interested in oxygen doping because it significantly p dopes the buffer layer (and SEG) and it is the only atmospheric gas that is stably absorbed on the buffer layer as was demonstrated by Turmaud [17]; annealing at 400 C in vacuum is required to desorb the oxygen (see also Sup. Mat.).

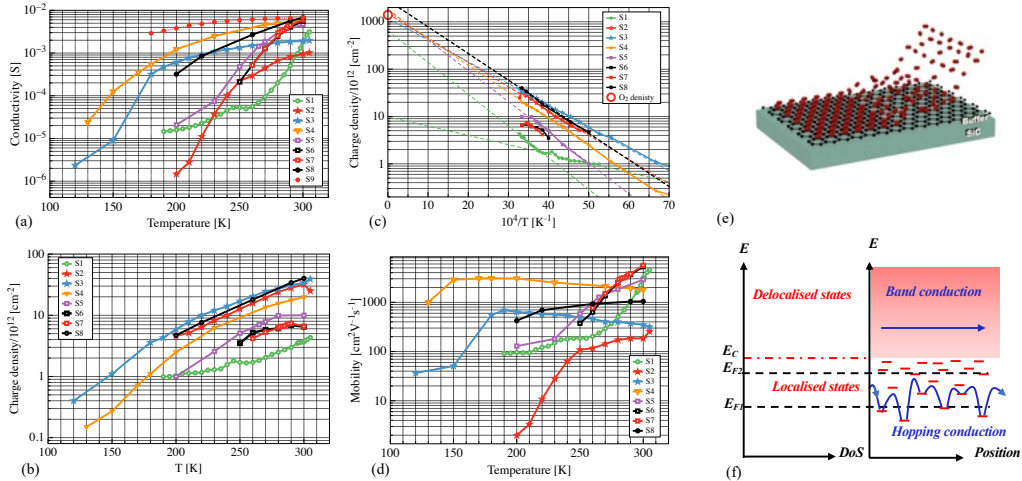

**Fig. 3.** Transport properties of SEG Hall bars, p-doped with oxygen at room temperature. (a) Conductivities are small at low temperatures and increase to up to about 5 x10$^{-3}$ S (= 200 Ω per square) at room temperature. This increase is attributed to increasing ionization of an adsorbed monolayer O$_2$ on the surface combined with sample dependent densities of impurity states. S9 was produced by draping 4 gold leaf contacts over an unprocessed ribbon. (b) The charge densities increase with increasing temperatures as can best be understood by in an Arrhenius plot, $n$ versus 1/T (d), which shows uniform slopes consistent with an activation energy of 120 meV which is attributed to the anionization of strongly physisorbed O$_2$ on SEG. Slope variations are due to variations of the dielectric properties, probably caused by surface contamination. Linear extrapolations to 1/T=0 extrapolate to about 1500 10$^{12}$ cm$^{-2}$ which is close to the estimated density of a monolayer of O$_2$ (large red circle). The small low temperature slope of S3 is consistent a 60% coverage of residual photo resist with a 10 meV activation energy that slightly p dopes the SEG. (c) The Hall mobilities show generally increasing mobilities with increasing temperature with mobilities ranging from 2-5500 cm$^2$V$^{-1}$s$^{-1}$. The mobility generally increases with increasing temperature dependence due to the increase in charge density as explained below. (e) Schematic diagram of the ionization process of an adsorbed oxygen monolayer. (f) The transport properties of doped SEG can be explained in terms of the transition from low mobility hopping transport via localized states in the band gap, to high mobility band transport, explained here in terms of electron transport (hole transport is formally equivalent). The process is essentially identical to transport in semiconductors with defects that produce localized impurity states in the band gap. As observed in CCS produced buffer layer at low charge densities and temperatures transport, the Fermi level is in the bandgap ($E_{F1}$) and transport is dominated by hopping from localized state to localized state resulting in low mobilities. The charge density increases with increasing temperature as shown in (b), causing the Fermi level ultimately to increase above the conduction band edge ($E_C$) so that the transport transitions to high mobility band transport. Consequently, the transition charge density, (and hence the transition temperature), depends on the defect density which about 0.27 x 10$^{11}$ for S4, 4.3 x 10$^{12}$ cm$^{-2}$ for S3, and 17 x 10$^{12}$ cm$^{-2}$ for S2. Details for the various samples are explained in the text.

The measurements were performed in a cryostat supplied with a superconducting magnet at temperatures ranging from 100 K to 300 K. At each temperature the magnet was swept from B=-3 T to +3 T for Hall measurements to determine the charge densities. Conductivities σ were determined from 4-point measurements and the Hall mobilities µ were determined from σ=neµ where *e* is the electric charge. Note that in 2D materials resistivities are expressed in Ohms and conductivities in Siemens.

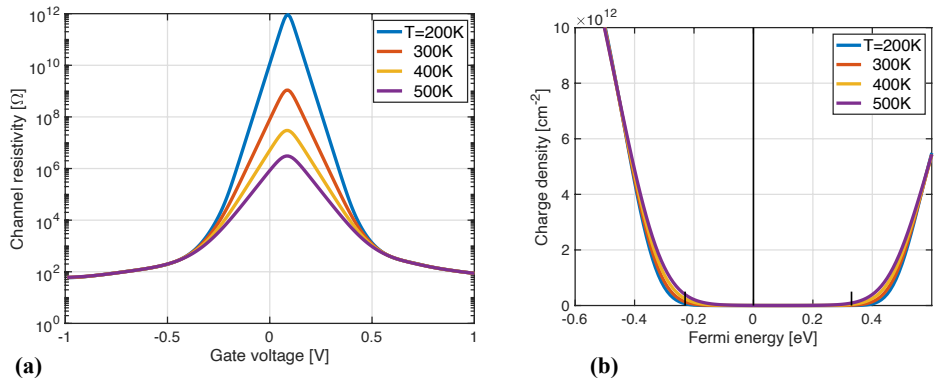

**Fig. 4** (a) Predicted SEG channel resistivity using the calculated density of states, with a SEG mobility of 4000 cm²V⁻¹s⁻¹ assuming an ideal dielectric, which predicts a room temperature on-to-off ratio that exceeds $10^6$. (**b**) Charge density versus $E_F$. The turn-on voltages for the N and P branches at T=300 K are predicted to be +0.34V and -0.23 V respectively.

The conductivities of the samples (Fig. 3a) all show a monotonic increase with increasing temperature. The room temperature conductivities range from 1 to 8 x $10^{-3}$ S corresponding to resistivities ρ from 125 Ω to 330 Ω. The low temperature values are up to a factor of 1000 smaller. Charge densities (Fig. 2b) range from ≈0. 2 x $10^{12}$cm⁻²- 40 x $10^{12}$cm⁻². The STS measurements (Fig. 2a) show that SEG is intrinsically charge neutral, so that the charging is caused by environmental gasses (including trace volatile organic compounds) and by residual resist from lithographic processing. [30]. The mobility (Fig. 2c) generally increases with increasing temperature tending to saturate at higher temperatures. The maximum measured mobility is 5500 cm²V⁻¹s⁻¹. The room temperature SEG conductivities, charge densities and mobilities are all within ranges that are typical for epigraphene. However, the temperature dependences are reminiscent of a doped semiconductor with deep acceptor states as elaborated below.

A semi log plot of the charge densities plotted as a function of $10^4$/ T Fig. (3d) clearly shows Arrhenius behavior. Note that the charge density n of p-type semiconductor with a doping density of $N_0$ is given by [31]

$$n = \frac{N_0}{1+g\times\exp\left(\frac{E_A-E_F}{kT}\right)} \qquad \text{Eq. 1a}$$

where k is Boltzmann's constant, $E_F$ is the Fermi level $E_A$ is the acceptor energy level, g is the acceptor degeneracy. For a deep level with g=1 at low temperatures ($E_A-E_F \gg kT$),

$$n \approx N_0 \exp\left(-\frac{\Delta E}{kT}\right) \qquad \text{Eq. 1b.}$$

Eq. 1 is derived for the ionization of deep acceptor states in a semiconductor [31], but the thermodynamics is similar for the ionization of neutral molecules on a 2D surface. From Fig. 3d we find that $\Delta E= 0.12 \pm 0.02$ eV. K We also find that $N_0 \approx 1.500 \times 10^{15}$ cm$^{-2}$ which is close to the estimated O$_2$ density of a O$_2$ monolayer (i.e. $1400 \times 10^{15}$ cm$^{-2}$ red circle, Fig. 3 [32]). These observations support the assumption that the p-doping is due to an approximately complete oxygen monolayer as can be expected. Variations in $\Delta E$ and $N_0$ are probably caused by partial coverages of trace environmental volatile aromatic molecules that are readily absorbed on graphitic materials reducing the oxygen coverage and affecting its ionization energy [33]. The observed two slopes in sample S1(Fig. 3d) are likely due to a significant ($\approx 60\%$) coverage of residual resist material which causes the p doping of SEG [34] at the low temperatures due to its small $\Delta E=0.035$ eV as determined from the slope.

The mobilities of samples S2, S3 and S4 show a steep rise followed by a plateau at transition temperatures $T_{tr}$ 250 K, 190 K, and 150 K respectively. The corresponding transition charge densities $n_{tr}$ are $17 \times 10^{12}$ cm$^{-2}$, $4.3 \times 10^{12}$ cm$^{-2}$ and $0.27 \times 10^{12}$ cm$^{-2}$. The mobilities at low charge densities result from localized defect states in the band gap [35] (Figs. 2e,f). The localized states are filled as the charge density is increased after which the transport transitions to high mobility band transport so that $n_{tr}$ is a measure of the defect state density. This process has been observed in 2D semiconductors [35] and it contributes to the subthreshold rise in thin film transistors[36]. In CCS produced buffer layers, the transport in the bandgap has been identified as variable range hopping [17] (Sup. Mat).

S4 was produced with a shadow mask which explains its small defect density. S9 (Fig. 3a) was produced by draping 4 gold leaf contacts over an unprocessed ribbon and that its conductivity is large and like S4, however it was not in a Hall bar configuration so that its charge density and mobility could not be measured. S2 was produced using O$_2$ and UV exposure can explain its large defect density and relatively low mobility. The decreasing mobility with increasing charge density in the post threshold plateau of S2,S3, and S4 is reminiscent of charge impurity scattering in graphene, [37]. S5,S6,S7 have large mobilities that increase with increasing temperature and a charge density that appears to saturate which we currently do not understand.

From the measured semiconducting and the density of states, we can predict the response of a field effect transistor, Fig. 4. The channel conductivity can be expressed as $\sigma(V_g) = n_e e \mu_e + n_h e \mu_h$ where $n_e$ is the electron density and $n_p$ is the hole density[38]:

$$n_{e,h}(E_F, T) = \pm \int_{E_F}^{\pm\infty} D(\varepsilon) F_{e,h}(E_F, T, \varepsilon) d\varepsilon \qquad (\text{Eq. 3})$$

Here $F_{e,h}$ are the electron and hole Fermi functions and density of states $D(\varepsilon)$ is from Fig. (2b) (red dashed line). With $\mu = 4000$ cm$^2$V$^{-1}$s$^{-1}$ an on-to-off ratio of $10^6$ (Fig. 4a) and a subthreshold slope (SS) of 60 mV/decade (Fig. 4b) which are amply sufficient for digital electronics [39]. An actual SEG field effect transistor is shown in Methods (Fig. 12) where the on-to-off ratio is already $10^4$, but its field effect mobility $\mu_{FET}$ is only 22 cm$^2$V$^{-1}$s$^{-1}$ primarily due to disorder caused by the dielectric and large Schottky barriers at the contacts.

**Conclusion and perspective**

The singular focus of epigraphene nanoelectronics research, even predating mainstream graphene research [40], was to develop a 2D nanoelectronics platform to succeed silicon electronics[5]. Graphene's lack of a band gap was universally considered to be the major hurdle towards this goal [39]. Here we have demonstrated that a well crystallized epigraphene buffer layer is an excellent 2D semiconductor with a 0.6 eV band gap and with room temperature mobilities that greatly surpass all current 2D semiconductors. A prototype FET has an on-to-off ratio of $10^4$ which may reach $10^6$ in optimized devices.

Besides mobilities that already reach 5500 $cm^2V^{-1}s^{-1}$, SEG is grown on THz-compatible SiC substrates[41] which have become an increasingly important commercial semiconductor compatible with conventional microelectronics processing methods[42]. In addition, epigraphene can be nanopatterned which is not possible with graphene on other substrates due to pervasive edge disorder on the nanoscale . In contrast the epigraphene edges turn out to be phenomenal one-dimensional conductors[43]. SEG can be intercalated with a wide range of atoms and molecules to form a wide range of materials with useful novel electronic and magnetic properties [44].

Current work focuses on reliably producing macroscopic terraces [45] producing viable dielectrics that do not severely reduce the mobility[46, 47], managing the Schottky barriers, and developing schemes to produce integrated circuits.

In the Methods section (see also Sup Mat), we briefly touch several of these points, where we convert SEG into quasi free-standing graphene (QFSG) by intercalating hydrogen where seamless SEG/QFSG junctions are realized mitigating interconnect problems[48, 49]. In conclusion, SEG opens a new chapter in 2D nanoelectronics with a significant potential to become commercially viable in the not-too-distant future.

# METHODS

## Sample Production

Samples used for transport measurements were produced in a closed cylindrical high purity graphite crucible, 14 mm long and 10 mm in diameter with a cylindrical bore of 5.5 mm. The crucible is closed and supplied with a cap, that is perforated with a 1 mm hole (Fig.1a). The crucible is placed in a quartz tube and heated inductively. The temperatures are monitored and controlled with an optical pyrometer. The typical sandwich is composed of two 3.5x4.5 mm SiC chips, where the Si-face of the top chip (the seed chip) faces the C-face of the bottom chip (the source chip). The Si-face of the source chip (i.e. specifically not between the chips) may be coated with a polymer which most likely causes its temperature to slightly increase relative to the top chip.

The sandwich is placed in the crucible and annealed in three phases as shown in Fig. 1. In the first phase they are cleaned of surface contaminants in high vacuum at 900 C. This followed by a high temperature pre-growth annealing step at 1300 C, in 1 bar of ultrahigh purity Ar, which causes the surface of Si-face of the top chip to develop a regular stepped structure (Fig. 1a, middle). In the final annealing step at about 1600 C, the Si-face of the top chip develops large atomically flat terraces (Fig. 1a, right). that are covered with a buffer layer (see Fig. 2d).

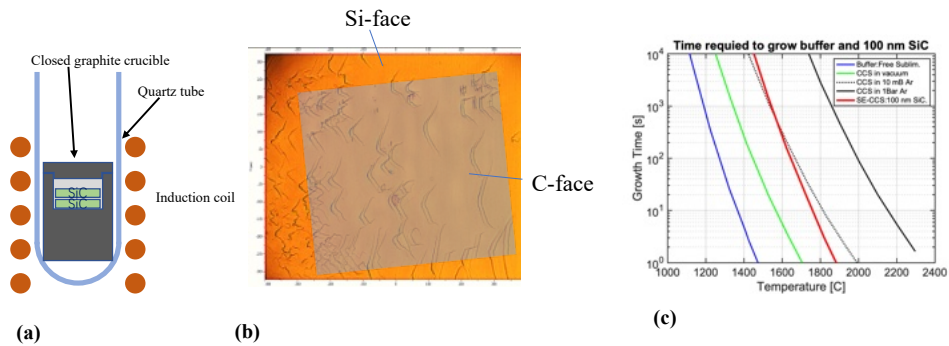

**Fig. 5 (a)** Vertical furnace with improved temperature gradient control. **(b)** Overlapped images of the surface of a Si-face seed chip and mirror image of the corresponding C-face source chip (slightly shifted) showing identical complementary topological features, i.e. material removed from the C-face is deposited directly above it on the Si-face which demonstrates the close interaction between the two chips. **(c)** Approximate times to grow a buffer layer and to grow 100 nm SiC from the source chip on the seed chip where the former is more than 10 C cooler than the latter

The graphitized polymer helps to establish the required temperature differential between the top and the bottom chip. However, entirely similar results have also been obtained without the polymer coating (i.e. Fig. 2e) in a vertical furnace (Fig 5a) in which the temperature gradient is controlled by the location of the sandwich in the crucible. A specific temperature difference within the relatively narrow annealing temperature window is found to be important. While nominally, the C face of the bottom chip is free of graphene, it becomes fully covered with graphene, with certain polymer coatings (Sup Mat). The Si face of the top chip always has a buffer layer covering step free terraces; however coverage is not always complete, especially in the middle of the chip. This may be due to a slightly increased Si vapor pressure there. The size of the terraces and buffer layer coverage vary considerably depending on temperatures and annealing times, polymer coating, SiC doping, SiC polytypes, and miscut size and direction. Current research focuses on optimizing these parameters, as well as developing alternative crucible designs in which the temperature gradient is better controlled, and with chips supplied with corrals that define growth areas [45].

Figure 5 (a) shows an alternative crucible design to study the effects of the temperature gradient where the gradient increases from the middle to the top of the crucible, as demonstrated in numerical simulations of the temperatures. This configuration is also used to demonstrate the mass transport from the hotter source chip to the cooler seed chip. For example, figure 5 (b) shows optical images of the surfaces of the Si-face seed chip superimposed with a mirror image of the C-face source chip, which clearly shows that their complementary nature, that is, material which is removed from the hotter C-face chip is deposited on the cooler Si-face chip. The C-face image has been shifted slightly for clarity. We also find that when the temperature gradient is inverted making the Si-face chip the seed source chip and the C-face chip the seed chip, then we find that the Si-face is covered with SEG but the (0001) terraces are small.

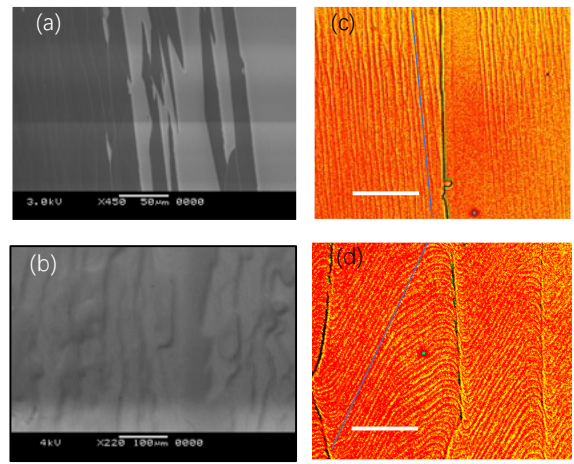

**Fig.** 6 Si-face source to C-face seed growth. (a) In this inverted geometry SEM images show SEG growth on the Si-face. (b) SEM of the C-face shows an irregular structure. (c) Contrast enhanced optical microscopy shows that the (0001) terraces on the Si-face are small but regular. (d) Contrast enhanced optical microscopy shows irregular step structure on the C-face which appears to be imposed by large steps (dark lines) the Si-face. Blue lines in (c) and (d) indicate the step directions of the unprocessed chips.

Figure 6 shows Si-face to Si-face growth of SEG. Here we see that the source chip (Fig. 6a) has relatively small regular terraces (≈ 10 μm) that are covered with SEG (darker areas) while the seed chip (Fig. 6b) no SEG is detected in Raman spectroscopy and the surface is irregular. Contrast enhanced optical images of the source chip (Fig. 6c) shows a regular array of parallel substrate steps whereas the step structure is highly distorted on the seed chip. Note that the large steps (dark lines) of the source chip are mirrored on the seed chip, which suggests that the source chip surface morphology is imprinted on the seed chip. Hence, in contrast to the Si-face to C-face configuration, large terraces do not form on either chip, moreover SEG forms on the source chip rather than on the seed chip.

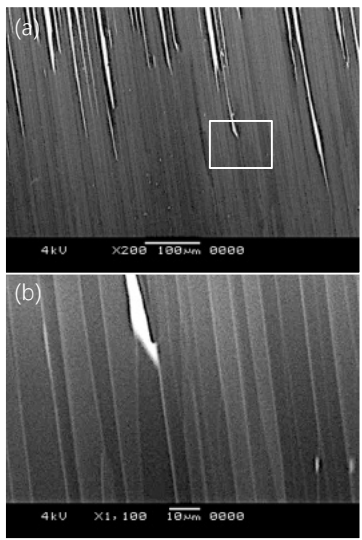

**Fig. 7** SEM image a single Si-face of a chip that is processed in a silicon saturated crucible. (a) the surface is largely covered with narrow (0001) terraces covered with SEG (darker areas). The white areas are bare SiC. (b) zoom in of boxed area.

We further find that SEG is stable in saturated silicon vapor. To show this we placed a single chip in a closed graphite crucible that has been saturated with silicon. As shown in Fig. 7a after annealing the Si-face is covered with SEG (darker areas) however the terraces are small.

We have also observed that in the standard C-face source to Si-face seed configuration, a Si is deposited on the C-face while SEG forms on the Si-face.

These observations lead us to conclude that the SEG coated (0001) face can is much more stable than the bare (0001) face and much more stable than all other SiC crystal faces. That is why when SiC is deposited from the C-face source- where the evaporation rates are known to be larger than from the Si-face- the newly deposited material favors SEG coated (0001) facets.

**Surface Characterization**

SEG and QFSG samples were characterized using an AFM (Park Systems NX10 with $50 \times 50$ μm scanning range) in non-contact mode for topology and contact mode for LFM measurements to identify graphene (Fig. 8). Microscopy was conducted using a scanning electron microscope (SEM, Hitachi SU3500, 15kV). Raman spectrometry was performed using a 532 nm laser with a spatial resolution of 1 μm (Fig.2d). STM and STS measurements were performed using a cryogenic scanning probe microscope (PanScan Freedom), Fig (2b,e). LEED measurements were made at the Georgia Tech Epigraphene Keck Lab (Fig. 2c). The representative SEM chip image (Fig. 2a) was made with a LEO 1530 FE-SEM (2 nm resolution), at 3kV in in-lens mode.

Figure 8 (top, middle) shows an SEM image of an atomically flat (0001) 300 μm wide terrace between two 100 nm high steps. An LFM scan was made over width of the terrace (white dotted line). No substrate steps were observed in that scan. In addition, 10 μm x 10 μm LFM maps were made at 3 locations (a,b,c) and no substrate steps or graphene patches were found, which confirms the SEM image and Raman spectroscopy of the terrace.

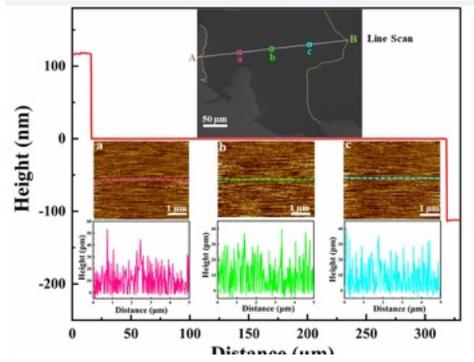

**Fig. 8** AFM measurement of an atomically flat SEG terrace between two ≈ 100 nm high substrate steps 300 μm apart. A single line scan, from spanning this distance, no SiC steps are detected, which are minimally 250 pm high. If there were substrate steps anywhere between the major steps, then this scan would have detected them. Topological 10 μm x 10 μm maps were made at three locations indicated, which did not detect any features larger than 50 pm, i.e., 5 times smaller than the minimal SiC substrate features, which verifies that SEG is atomically flat.

**Transport measurements**

Photoresist S1805 was spin-coated on the Si-face of the top chip, which is then photolithographically patterned using a direct laser writer (SVG-Micro 100). Alternatively, some samples were patterned using a shadow mask to test the importance of contamination introduced by the lithography process. The shadow masked areas show significantly smaller defect densities (i.e. sample S4, Fig. 3)

Contacts and gate electrodes were produced from 10 nm Cr and 30 nm Au e-beam (EB-500) deposited films followed by lift-off. Alumina dielectrics were produced by e-beam deposition of 2 nm Al that was oxidized in residual $O_2$, which served as the seed layer for the subsequent ALD process. Hall bar structures were typically 300 μm long and 30 μm wide.

Measurements were made on top gated samples and on oxygen doped samples. The latter are critical to get a high lower bound of the intrinsic mobility and other transport properties of SEG, which top gated samples cannot provide due to the severe scattering caused by the dielectric. We chose oxygen because it is safe, easy to use, stable on SEG[17], the charge densities can be varied over an order of magnitude, and easily removed by annealing in vacuum at moderate temperatures. Most importantly, it has been previously studied in a CCS produced buffer layer which allows a direct comparison with that system (see Sup Mat). Ammonia, (n-dopant) like several other graphene dopants, rapidly desorbs at room temperature (Sup Mat). It, and other dopants will be studied in future comprehensive investigations in rigorously controlled conditions.

Cryogenic transport measurements reported in the main text were made in an Oxford Instruments cryostat, Teslatron PT (300mK-305K) with a maximum magnetic field of 14T. Measurements were made using a Keithley 2450 SourceMeter, a Stanford Research SR560 voltage amplifier, a SR830 lock-in amplifier, a voltage amplifier and a DL Instruments 1211 current preamplifier. Four and 2-point measurements were made at temperatures ranging from 120 K to 305 K. For each measurement the temperature was stabilized, and the magnet was swept from -3 T to +3 T.

Measurements were also made in a cryogenic probe station (Lakeshore-Model TTPX) with a semiconductor device analyzer (Keysight-B1500A) for the FET measurements and for 2-and 4-point measurements on unprocessed natural SEG ribbons that were contacted using mechanically transferred prefabricated gold leaf strips (Fig. 3a sample S8). While the charge densities of these devices could not be determined, its properties are consistent with highly pure SEG ribbon when compared with the processed samples in Fig.3.

**Quasi freestanding graphene**

SEG was converted to quasi freestanding graphene (QFSG) by hydrogen intercalation [48] [49], which passivates the Si-bonds. Surface studies of QFSG (Fig. 9) show an essentially perfect graphene lattice structure as expected, and Raman maps confirm that the surface is free of SEG. Transport measurements on QFSG Hall bars (Fig.10) show that at room temperatures, mobilities, charge densities, and resistivities are remarkably like SEG. However, they are quite different at lower temperatures due to the band gap in the SEG samples. The high mobilities at room temperature is primarily due to the weak electron phonon interaction[18] and the low defect densities.

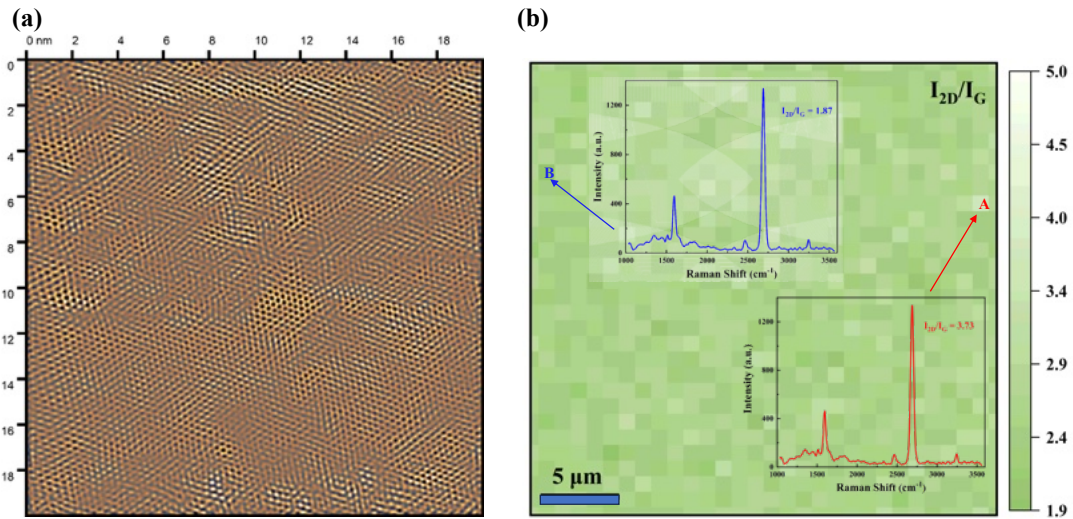

**Fig. 9** QFSG characterization. (a) Low temperature STM of a 20 μm by 20 μm area of QFSG produced by hydrogen intercalation shows that it is defect free. (b) Raman map of a 25 μm X 25 μm area shows that it is completely covered with graphene with no bare SiC or buffer. The arrow labeled A points to a region with a $I_{2D}/I_G$=3.73 (red scan) and the arrow labeled B points to a region with a $I_{2D}/I_G$=1.75 (red scan). Variations of this size are normal for graphene.

QFSG is also important for nanoelectronics since it facilitates a seamless contact between QFSG and SEG, which will be important for nanoelectronics since metallic nanoscale contact to SEG will be challenging. Figure 10 shows an example of such a junction and in Sup. Mat. we show transport measurements between graphene and CCS produced buffer layer. Also note that a wide range of materials can be intercalated under SEG which can be used to mitigate Schottky barriers between QFSG and SEG as well as providing interconnects in integrated nanostructures.

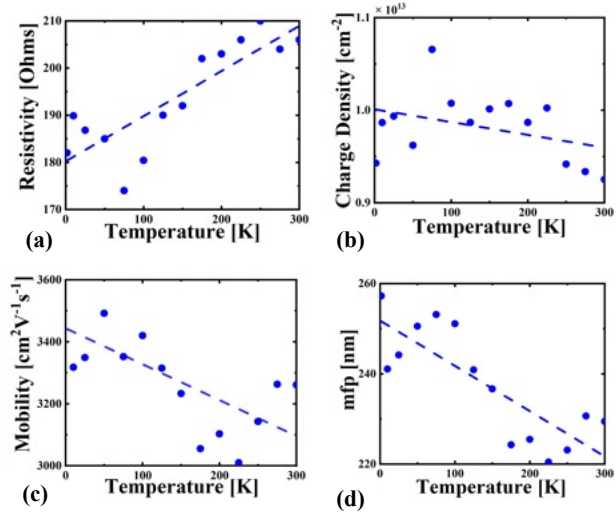

**Fig. 10** Transport measurements of a QFSG Hall bar. (a) Resistivity versus temperature, (b) Charge density versus temperature (c) Mobility versus temperature. (d) Mean free path versus temperature. Note the absence of a significant temperature dependence compared with SEG (Fig. 5). Also note that at room temperature, the charge densities and the mobilities are comparable to those of SEG.

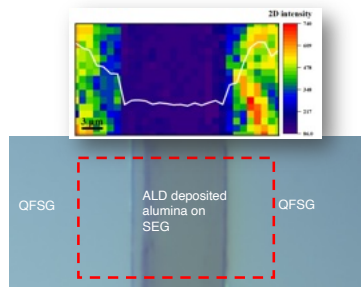

**Fig. 11** Example of seamless SEG/QFSG junctions, produced by depositing an $Al_2O_3$ strip, 80 μm wide, and intercalating hydrogen at 700 C.

**SEG field effect transistor**

The electrical properties of the SEG were measured by characterizing a fabricated top-gated SEG FET. Figure 12a shows a schematic drawing of the device. The transfer curves are plotted in figure 10b with $V_{ds}$ of 0, 1.0 and 2.0 V. The device shows ambipolar characteristics. As $V_{ds}$ increases, both $I_{on}$ and $I_{off}$ monotonically rise. As shown in Figure 12c, the device exhibits reasonable switching performance with an on/off ratio of ~$10^4$ at $V_{ds}$=1 V and $I_{on}$=15 nA. The threshold voltage ($V_{Th}$) is -0.21 V which is extracted by extrapolating the linear regime of the transfer curve to the gate voltage axis. The subthreshold swing (SS) calculated from SS=$dV_{gs}$/ $d(\log I_{ds})$ is ~155 mV·dec$^{-1}$Figure 12 (d) shows output curves of the device. There is a substantial barrier (SB), which is clear from the non-linear behavior of $I_{ds}$ at high $V_{ds}$ and the large contact resistances. Figure 12e extrapolates the linear rise of the output curve to the baseline, which closely corresponds to the band gap (Fig. 2e). Improving the metal contacts and the quality of dielectric layer will significantly enhance the device performance to approach the theoretical values calculated in Fig. 4.

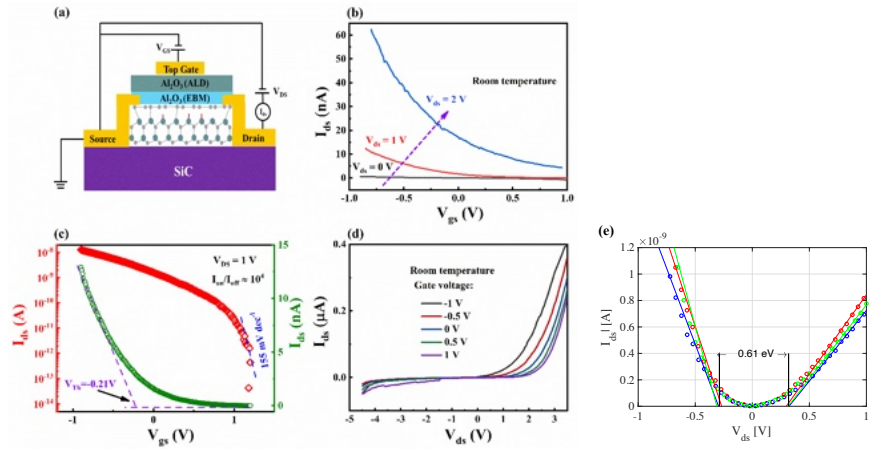

**Fig. 12** (a) Schematic of field effect transistor with SEG as channel. (b) Transfer characteristics ($I_{ds}$-$V_{gs}$) at bias voltage of 0, 1 and 2 V. (c) Transfer curve of the device at $V_{ds}$=1 V and corresponding logarithmic plot. (d) Output characteristic curves of the device. The field effect mobility is µFET=22 cm$^2$V$^{-1}$s$^{-1}$. The large reduction compared with the intrinsic SEC properties is caused by scattering from the dielectric and large contact Schottky barriers. (e) Extrapolation of the linear rise of the output curves correspond well with the band gap (Fig.2e)

# Supplementary Information

# SEG production

**A comprehensive account of the SEG face-to-face production method and analysis**

1. Face-to-face configurations

SEG growth was accomplished in a variety of chip configurations. In the following tables the top row gives the placement and the polytypes involved. Measurements Seed chip are given. For example, in the left the column of measurements in Table 1, the Seed is the Si face of an N doped 6H polytype chip, and the source is the C-face of a 4H polytype N doped chip. In the SEM micrographs, dark areas are SEG, and light areas are SiC.

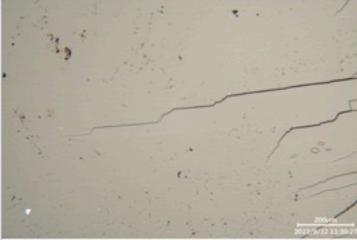

**Table 1**. Standard growths done with various SiC polytypes.

| Si-face Seed (shown) / Si-face Source | 6H / 4HN | 4HN / 6H | 4HN / 4HN | 6HN / 6HN |
|---|---|---|---|---|
| Summary | There are big steps, covered by Buffer in the middle | A small amount of large steps in the middle cover the Buffer | A small amount of large steps in the middle cover the Buffer | A small amount of coverage of large steps Buffer |
| Optical Microscope | 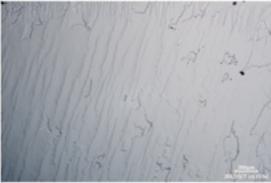 | 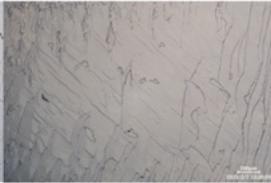 | 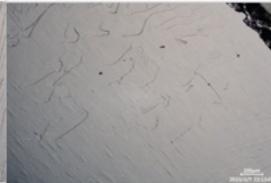 | 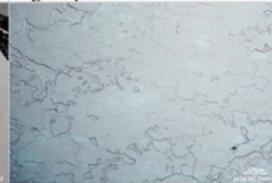 |
| Raman | 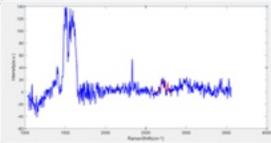 | 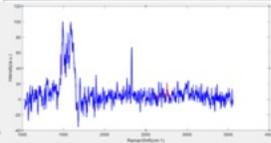 | 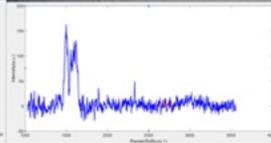 | 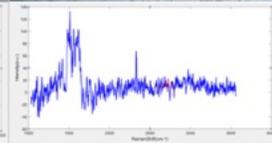 |
| SEM | 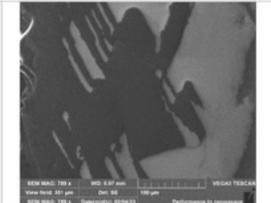 | 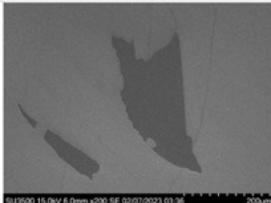 | 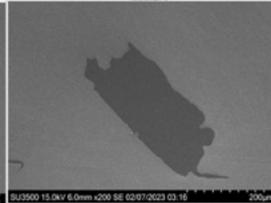 | 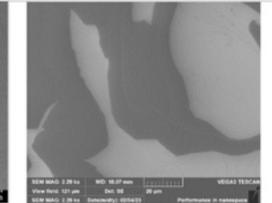 |
| AFM | 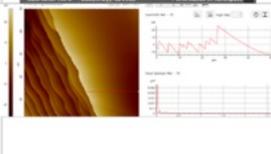 | 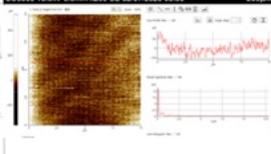 | 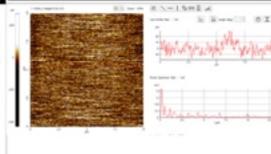 | 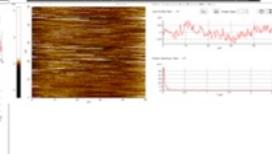 |

| Si-face Seed (shown) / Si-face Source | 4H / 4H | 4H / 6H | 6H / 6H | 6H / 6HN |
|---|---|---|---|---|
| Summary | A small amount of coverage of large steps Buffer | Few large steps on the edge, covering Buffer, no Buffer in middle, and spiral structure found | More coverage of large steps; buffer | Large steps, high buffer coverage |
| Optical Microscope | 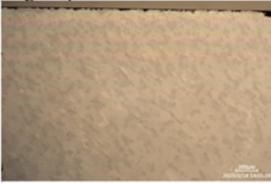 | 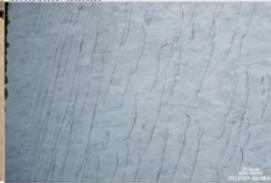 | 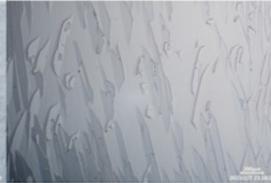 | 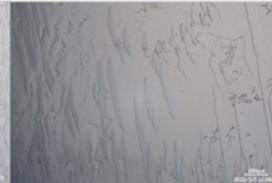 |
| Raman | 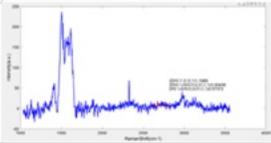 | 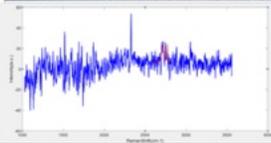 | 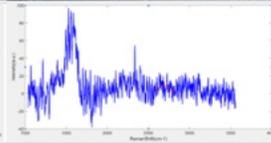 | 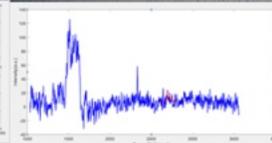 |
| SEM | 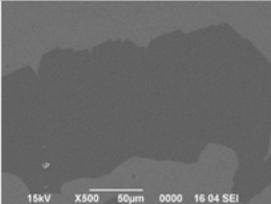 | 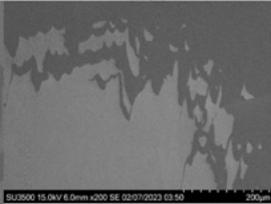 | 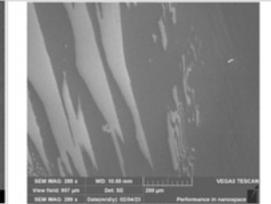 | 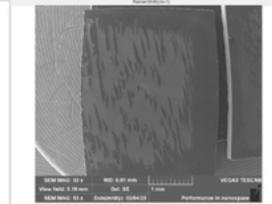 |
| AFM | 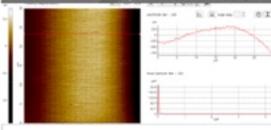 | 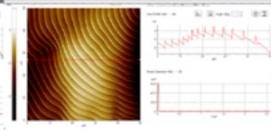 | 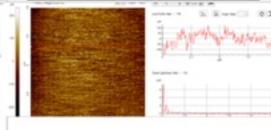 | 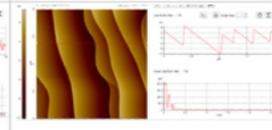 |

**Table 2 top; Table 3 bottom** 3. Si-face-to-Si-face growths done with various SiC polytypes.

| Polymer | SC1813 | SC1805 | PMMA – 495A4 | Amorphous Carbon |
|---|---|---|---|---|
| Summary | There are big steps, covered by Buffer in the middle | There are big steps, covered by Buffer in the middle | There are big steps, covered by Buffer in the middle | There are no large steps in the middle, and a small number of |
| Optical Microscope | 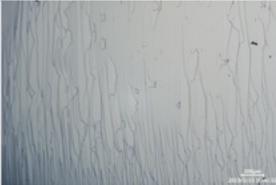 | 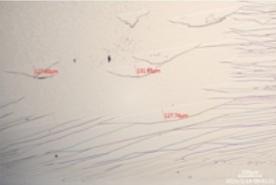 | 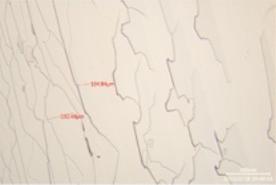 | 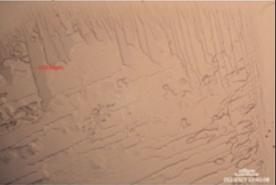 |
| Raman | 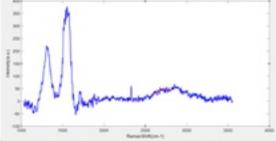 | 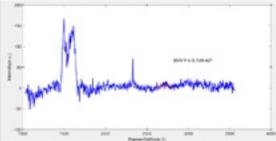 | 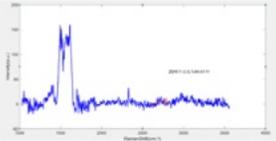 | 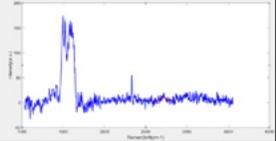 |
| SEM | 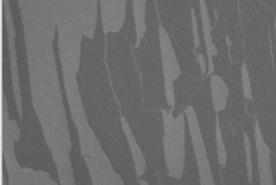 | 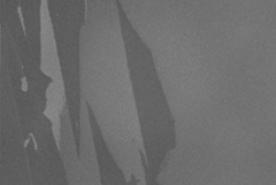 | 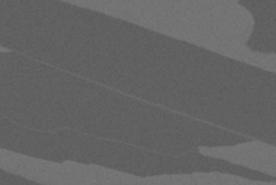 | 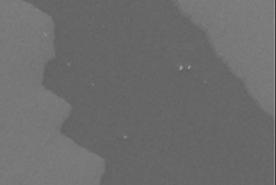 |
| AFM | 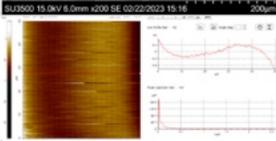 | 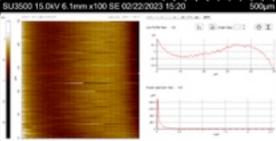 | 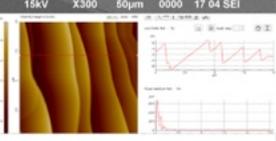 | 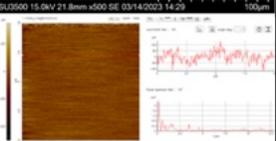 |

**Table 4.** Standard growths done with various polymers coating the silicon face of the source chip and the associated growth on the silicon face of the seed chip.

## 2. SEG growth

**High SEG covered chips.**

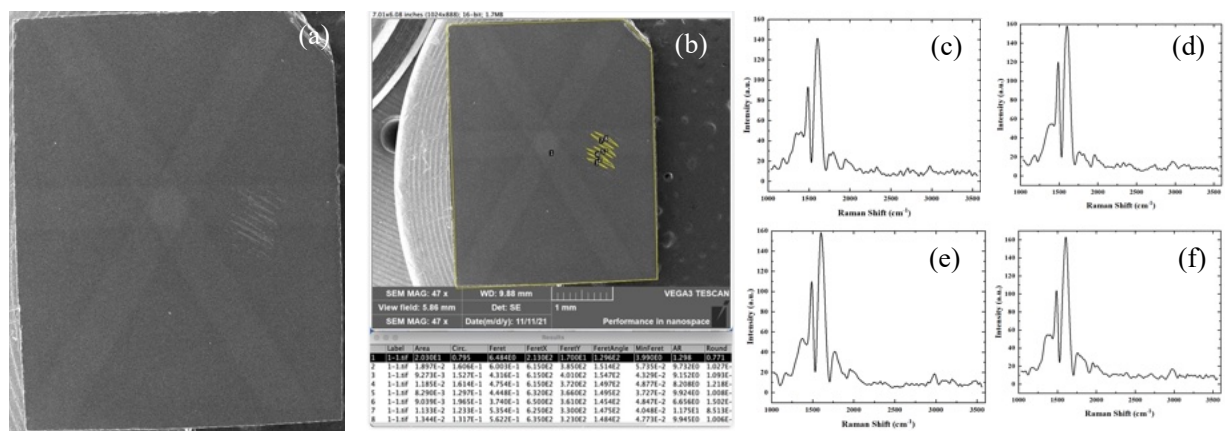

**Figure S1**. (a,b) SEM images of completely covered SEG samples with (c-f) sampled Raman spectra. The coverage rate of SEG was calculated using values at the bottom of (b) (20.3-0.019-0.009-0.0118-0.009-0.0113-0.01344)/20.3 = 99.63% coverage. The feature of this sample of that SEG patch is rather small.

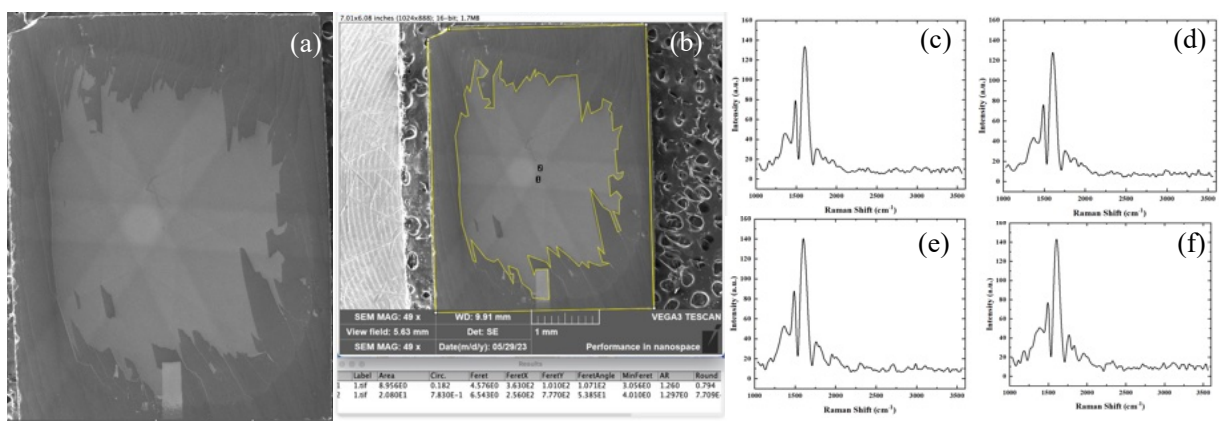

**Figure S2**. (a,b) SEM images highly covered SEG samples with (c-f) sampled Raman spectra. The coverage rate of SEG was calculated using values at the bottom of (b) (20.8-8.95)/20.8=56.97% percent coverage. The feature is that all SEG patch is on the order of several hundred micrometer. Dark areas are SEG, light areas are SiC.

**Confirmation of uniform terrace coverage**

Here, we investigate the amount of buffer layer coverage on a single terrace. We expect the buffer layer to be either undergrown (indicated by patches of silicon carbide on the surface), overgrown (indicated by patches of graphene on the surface), or optimally grown. Scanning electron microscopy reveals consistent contrast across the entire terrace surface at large scale (Fig. S3a) or small scale (Fig. S3b). If silicon carbide was present on the terrace, we would see bright features, and if patches of graphene were present, dark features would be readily apparent. Their differences in contrast can be seen in figure S3c. The steps themselves are silicon carbide, as shown in figure S3b, indicating that growth starts from the step edges.

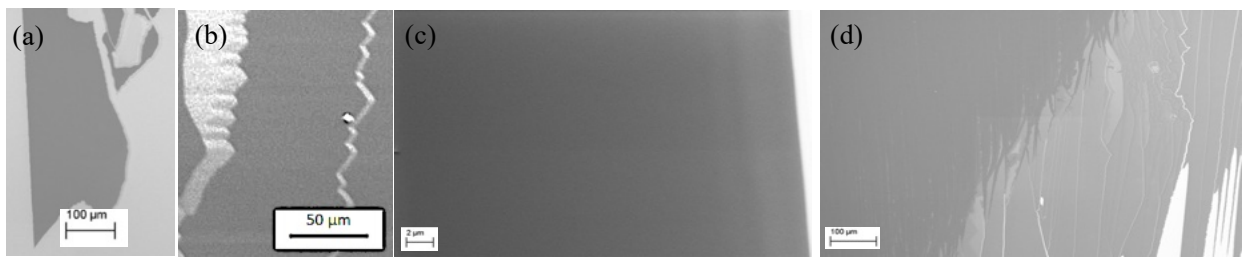

**Figure S3**. SEM images of buffer terraces showing uniform, defect-free coverage. (a,b) Example of a large terrace with uniform coverage with a (c) zoomed in image (3KV, WD = 13.1mm). Changes in contrast occur due to the edges of the terrace. (d) Boundary of chip overlap showing changes in contrast from dark grey (graphene) to light grey (buffer) to white (silicon carbide). Darker colors in the images correspond to higher conductivities.

Lateral force microcopy further supports the claim of continuous, uninterrupted growth. On measurements along a terrace edge (Fig. S4a) and center (Fig. S4b), the only substantial changes in frictional force occur along the step edge due to the presence of large topographical changes. Due to silicon carbide and the buffer layer's relatively high friction significant drops in lateral force would indicate the presence of graphene on the surface, owing to its extremely low frictional force.

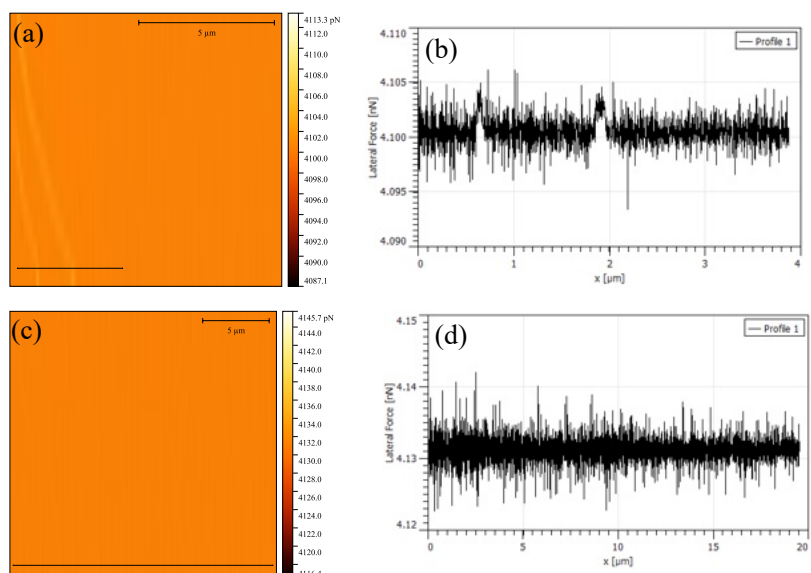

**Figure S4**. LFM Measurements of buffer terraces and linescans (a,b) close to step edge and (c,d) away from a step

Uniform coverage is further supported in the main text by the Raman mapping (Fig 2c) of a 50x50µm area with little variation, and LEED (Fig 2b) showing no graphene spots.

## Terrace Formation

The formation of large-scale terraces of the buffer layer are due to a local quasi-equilibrium between the silicon face of the seed chip and the carbon face of the source chip. Although some material escapes from between two chips, the growth is a result of a local material transfer and causes the two faces (seed silicon-face and source carbon-face) to create almost mirrored images of each other.

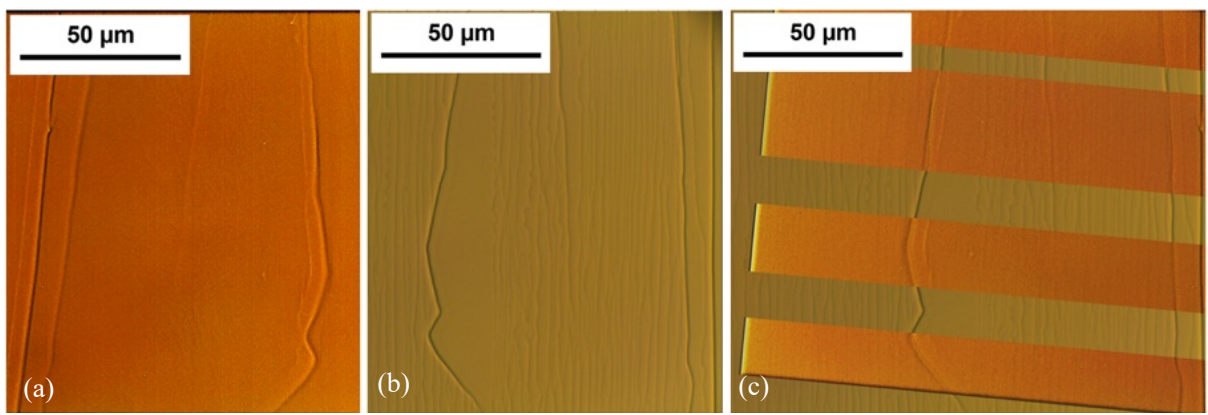

**Figure S5**. Optical images of a single terrace interface. (a) is the C Face feature (b) the Si-face feature, and (c) shows the C-face overlaid on the Si-face to show the matching. To make the C-Face steps clear, the image contrast and sharpness were increased, and sections were cutout in the overlay to make it easier to see the matching. Note this sample was grown with reversed temperature gradient in a horizontal furnace, but the effect is the same in the standard growths (Fig 5).

In the vertical furnace, a lower temperature is needed to establish a gradient between the two samples. The growth flows "downhill," since the carbon-face source sublimates silicon at a lower temperature. As the terraces grow, the seed silicon-face needs to build new silicon carbide on top of the existing steps to maintain an atomically flat terrace. The most readily available source of silicon carbide being the hotter source chip directly below the terrace. The trailing edge of the terrace starts out flat, since the amount of silicon carbide required to build the initial steps is low, but eventually steps up to a peak at the leading edge of the terrace and steps down very shortly after. The step-up on the carbon face terrace correspond to a miscut angle of about 0.6%, which is closer to the miscut angle of the seed chip than its own original miscut at 0.1%. This could be an effect of the different miscuts. The carbon-face chips are much flatter, so if the distance between the two surfaces stays relatively even, it must grow steps opposite its normal step direction.

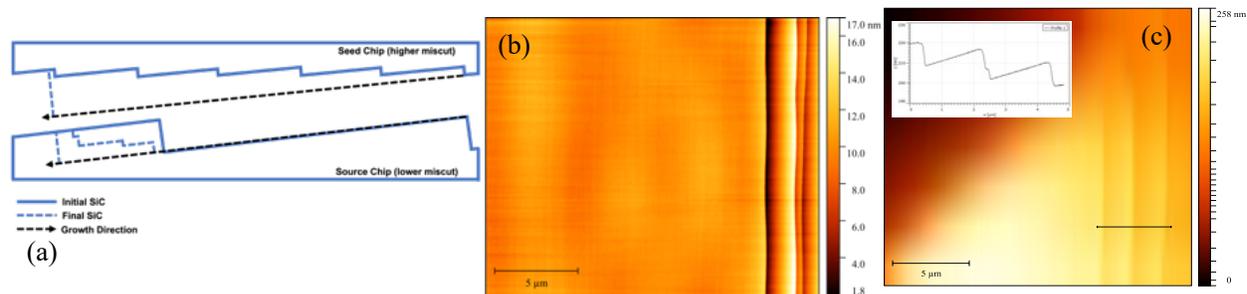

**Figure S6**. (a) Schematic of terrace formation at the interface of the chips with AFM images at the (b) start and (c) end of a mirrored terrace on the carbon face of the source chip. The inset shows a profile of the steps increasing as you get closer to the terrace edge.

The growth of a buffer layer on the silicon face of the seed chip implies there is a net silicon loss at the interface– either to the source chip or to the system. If the silicon condensed on the carbon face of the source chip in the form of polysilicon or silicane, conventional surface characterization methods would detect its presence. Either through a honeycomb lattice structure in the case of silicane or a thick oxide layer in the case of polysilicon. LEED (Fig S7a) shows no additional crystal structures other than silicon carbide's hexagonal lattice. Raman spectroscopy (Fig S7b) shows no difference between on and off terraces, and energy dispersive x-ray spectroscopy (Fig S7c) fail to detect significant oxides on the surface of the carbon face. Even scratching the surface with stainless steel shows no change in the surface morphology. This leads one to believe that silicon is simply lost to outside of the terrace interface, which is directly supported by larger terrace growth near the edge of the sample (Fig 2e) where lower surrounding silicon vapor pressures assist the silicon in diffusing out.

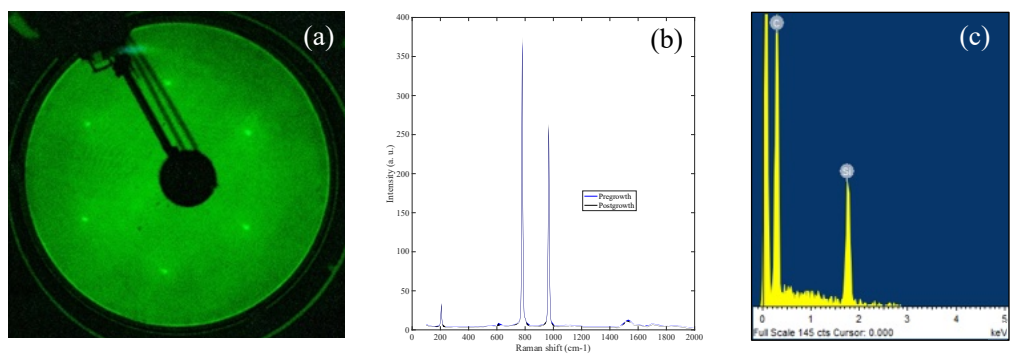

**Figure S7**. Characterization of the source chip carbon face. (a) LEED at 125V of a chip immediately after the growth process showing only SiC spots. (b) Raman spectra taken pre and postgrowth, and (c) EDS of a carbon face terrace at 3kV. Oxygen would be expected at 0.525kV, but only background signal is present.

We now analyze the various factors in this growth process.

*Temperature Gradients*

The vertical temperature gradient can be reversed and still grow the buffer. With the 'seed' chip silicon-face up, underneath the 'source' chip (carbon-face down, facing the 'seed' silicon-face), in a horizontal crucible with no polymer (likely a small temperature gradient), buffer terraces were grown at 1700C. The same was done in the vertical crucible with similar growth around 1650C. Terraces confirmed to have buffer in Raman, no graphene found. Outside the terraces, only plain silicon carbide was found in Raman. This sample also displays the mirroring behavior of the two surfaces.

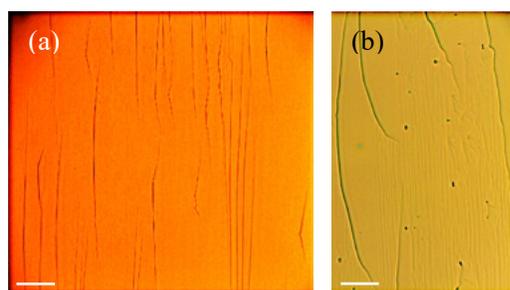

**Fig S8**. Optical images of samples grown with reversed temperature gradient in (a) horizontal furnace and (b) vertical furnace. Scale bars are 100μm and 20μm for (a) and (b) respectively.

Larger temperature gradients were trialed in the vertical furnace, but all attempts led to seeded growth of silicon carbide on the top sample.

*Temperatures and times*

The terrace size is primarily affected by temperature while the number of terraces (and therefore the overall buffer coverage) is affected by time. Note the total material flow required to form a terrace is quadratic in the terrace width (perpendicular to the natural step flow). The terraces size appears to saturate, while new terraces to form. Ultimately terraces cover nearly the entire chip, however they do not increase much in size.

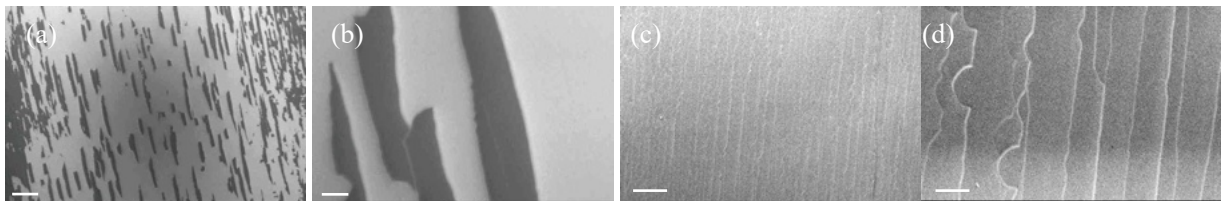

**Figure S9**. (a,b) show a sample grown for 1 hour at 1700C in horizontal furnace, with isolated buffer terraces. (c,d) show a sample grown for 8 hours at 1700C. (c) shows nearly complete buffer coverage. Both samples have terraces of about 20 microns. Scale bars are 200µm, 20µm, 100µm, and 20µm for (a), (b), (c), and (d) respectively.

After 8-hour growth we observed nearly complete buffer coverage however graphene was not observed anywhere in the chip interior. This agrees with quasi equilibrium buffer growth.

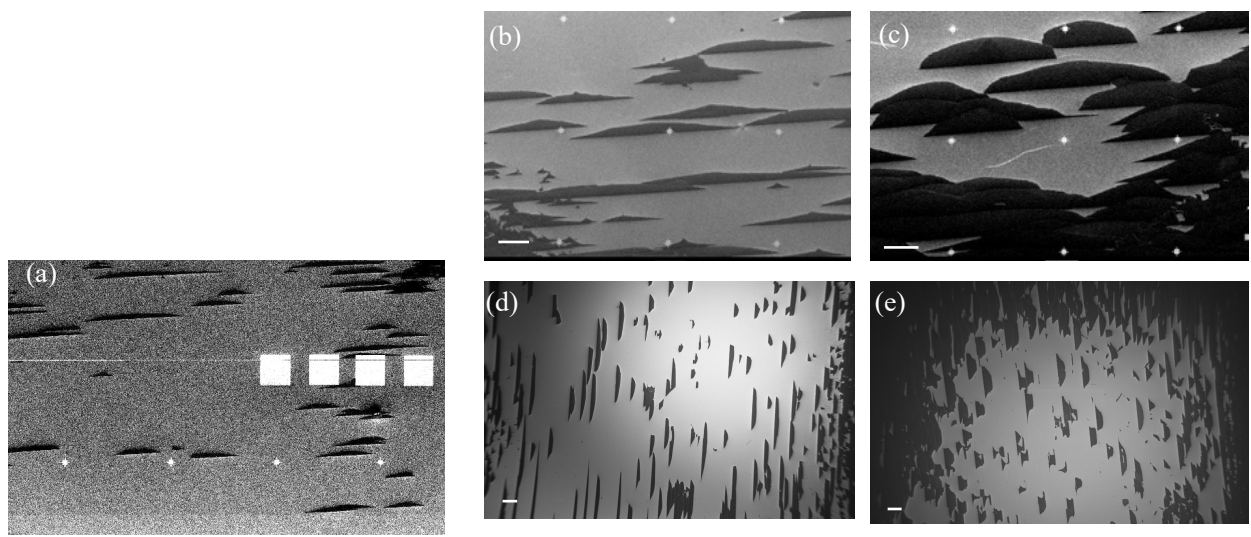

**Figure S10.** SEM Images of various sample growths in the vertical furnace. (a) Reference sample with low terrace growth and small size. Terrace density increases when time is increased by (b) one hour and (c) two hours. Terrace size increases with temperature when increased by (d) 25C and (e) 75C.

*Si Saturation*
Several growths were done in a silicon saturated crucible created by melting a silicon piece into the graphite at 1500C. The terrace formation and buffer growth seem unaffected, the chips in figure S9 were grown with silicon saturation. The silicon face of the source ship, in these growths, ends with either buffer layer or bare silicon carbide. If it's silicon carbide, that implies any silicon evaporated to form the buffer layer must stay within the chip interface. If it's buffer, that implies silicon saturation can be used to form uniform buffer layers quite easily, since it completely suppresses graphitization.

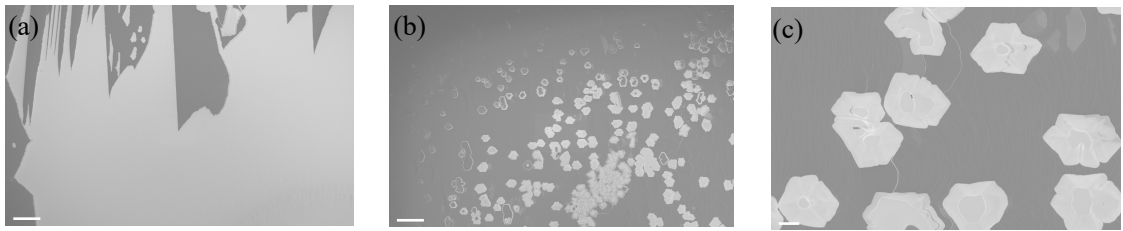

**Figure S11.** SEM images of growths completed in the vertical crucible with (a) sealed cap and (b,c) cap with 1mm hole. Scale bars are 100µm for (a) and (b) and 10µm for (c)

To lower the silicon vapor pressure outside of the chip interface, a hole was drilled in the cap of the vertical furnace crucible. What started as uniform terrace growth quickly turned into seeded silicon carbide growth. The original steps of the seed chip quickly formed a buffer layer halting step flow (Fig S11c). Material escapes too quickly from the source chip and deposits directly on the seed chip.

*Miscut Angle*
The wafer dicing cuts the surface at a slight angle to the (0001) plane, creating a somewhat even stepped surface. The miscut angle directly affects the growth of step free terraces since the higher the miscut, the more silicon carbide must flow to get the same terrace width. On the wafer with a 0.1% miscut (AI), large terraces were formed at 1600C. On the 0.4% miscut wafer (AJ), large terraces only started to form above 1650C, with best results at 1700-1735C.

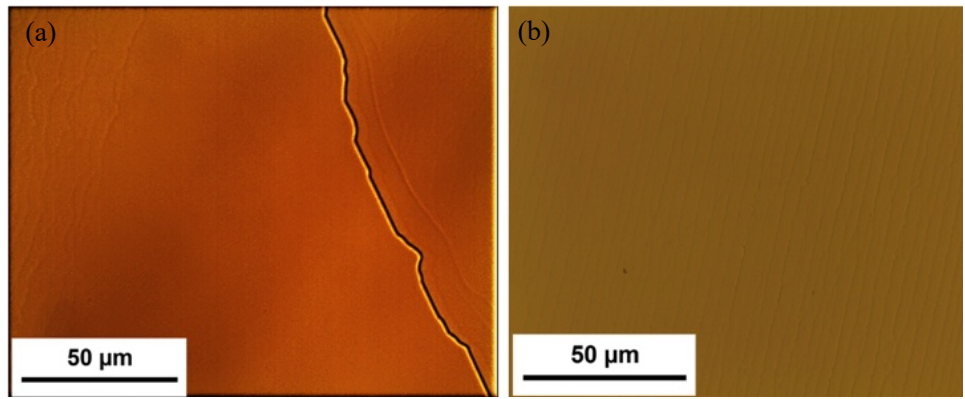

**Fig S12.** Optical images of chips grown face to face at 1600C for one hour. (a) chip is from a wafer with 0.1% miscut and (b) is from a wafer with 0.4% miscut.

*SEG growth in various chip configurations*
SEG growth using the face-to-face method is very flexible and allows a large range of configurations, involving semi-insulating SiC, N doped SiC, and the polytypes 4H-SiC and 6H-SiC. Moreover, whereas the devices measured here used the originally discovered configuration with and C-face source chip and a Si-face seed chip using semi-insulating SiC, similar results are found in other combinations, with adjusted temperatures. In the following pages we present a summary of the results of 15 such combinations with Optical images, SEM images, Raman spectroscopy, and AFM maps of the seed chips. Essentially similar results are found in each case.

3. **Properties of the buffer layer produced by the conventional CCS method** [1].

*Edited, selected excerpts from Jean Phillipe's PhD thesis with the authors permission. Edits indicated by []. Here we contrast results from CCS buffer with SEG, clearly showing disorder in the former, leading to variable range hopping transport and low mobility SEG. Here we also demonstrate a seamless graphene/buffer/graphene device.*

# Variable Range Hopping Conduction in the Epitaxial Graphene Buffer Layer on SiC(0001)

Jean-Philippe Turmaud

Georgia Tech. June 2019

**Face-to-Face growth**

Step flow and step bunching naturally occur on the Si-face of SiC during graphene growth [2]. To gain a control of the step flow, amorphous carbon corrals have been proposed as a mean to pin the SiC surface during growth, forcing the steps to bunch at those carbon barriers and creating large flat terraces during graphene or buffer layer growth [3] . In a more general manner, promoting a large step bunching during or before graphene growth allows to produce large step free buffer layer areas. We adopted a method along those lines, referred to as "face-to-face" annealing. It consists of stacking two (or more) pieces of SiC in a closed (no leak hole) graphite crucible, with their (0001) surfaces facing each other. Due to this confinement and the exchange of Si atoms from one surface to the other, graphene growth is inhibited up to at least $1700°$ in an argon background pressure above 1 atm. Step flow and step bunching on the two (0001) surfaces are then largely promoted: Fig. 4.2 shows the result of such an annealing for two hours for (a) a $4°$ off-axis SiC wafer and (b) an on-axis wafer. The smaller the miscut angle is, the larger the terraces can be obtained.

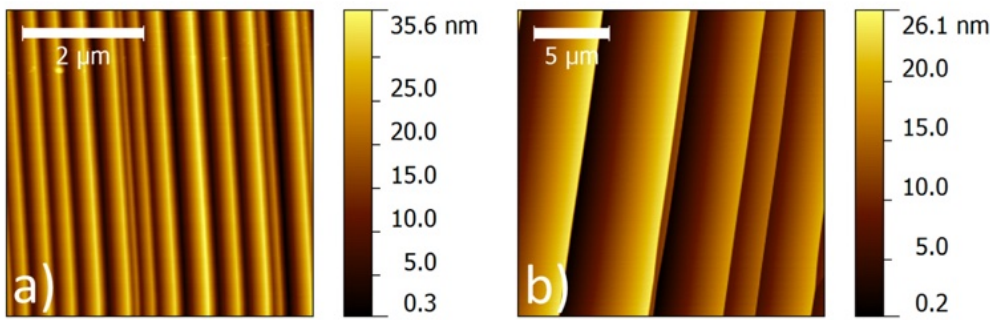

**Figure 4.2**: NC-AFM topography images of SiC surface after face-to-face annealing for a) $4°$ off-axis 4H SiC and b) on-axis 4H SiC.

**CCS production of the buffer layer**

[The buffer layer sample is grown at $1350°$ C for 30 minutes in a CCS furnace, like Fig. 1a (main text), where Si evaporates from the SiC surface, so that the surface becomes carbon rich and forms a buffer layer.

In contrast to SEG growth, only a single SiC chip is used after the face-to-face treatment above, to produce 5 μm scale terraces.]

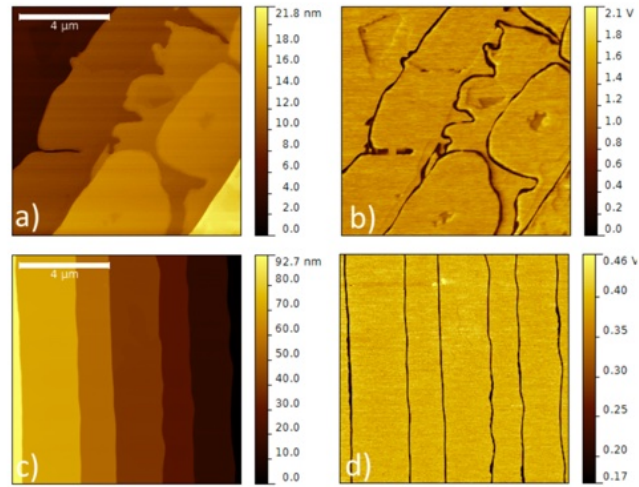

**Figure 4.3**: Fig. 4.3 a) and b) shows the meandering of steps due to the buffer growth process, and "canyons" forming inside the terraces for steps averaging around 5 nm high. The surface structure has a much higher probability to be preserved when the steps are higher than 10 nm, as shown in Fig. 4.3 c) and d).

**STM and STS of CCS produced buffer layer.**

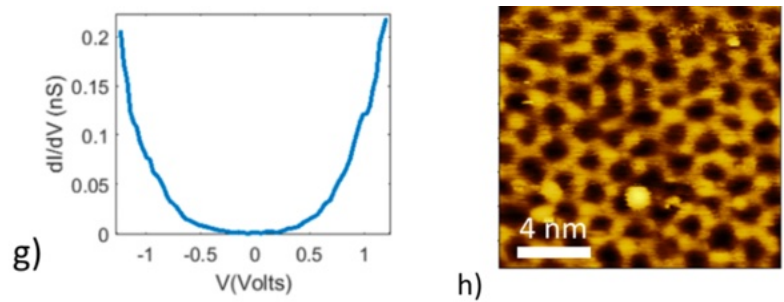

**Figure 3.2**: STS and STM images taken on the [CCS produced] buffer layer…. g) STS and h) STM tunneling voltage $V_t$=1V and current $I_t$=100pA .

[Note the disorder in the STM image compare with Fig. 2a (Main text) and the large density of state in the band gap measured by STS. Similar results have been reported on free sublimation produced buffer layer by others [4] [5] ]

**Graphene to buffer layer to graphene device**

Single wall graphene ribbons (SWGNR) form on the sidewall of natural SiC steps [and etched trenches] [6]. These nanoribbons can be grown in a controlled manner and utilized as electrical leads to inject current in the buffer layer, as SWGNR are seamlessly connected to the buffer layer on top of the trenches Fig. 4.1 shows a proposed geometry for a top gated buffer layer device, where the channel is made of the buffer layer grown on top of an etched trench. The seamlessly connected SWGNRs act as source and drain for the device. Alternatively, devices can be built with Pd/Au contacts directly deposited on a SiC step free buffer layer area.

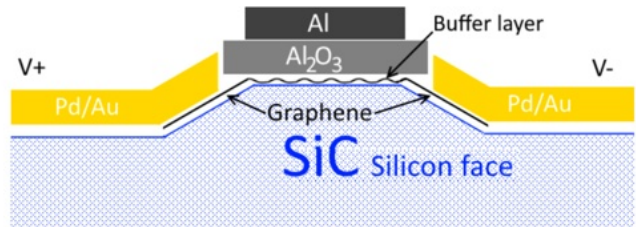

**Figure 4.1**: SWGNRs and buffer layer seamlessly connected in a top gated device geometry.

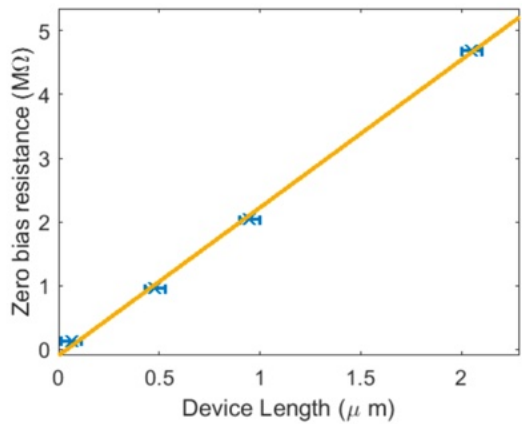

**Figure 4.8:** Zero bias resistance of a series of graphene contacted buffer layer devices (Fig. 4.1) plotted versus the length of the devices.

[2 point resistance measurements of several devices with 4μm wide buffer layer channels and several length show that] the resistance is clearly an increasing function of the channel length, and an estimate of the contact resistance can be extracted from the intercept at zero length. In both cases, it is virtually negligible compared to the 10 to 20 MΩ bulk resistivity (extracted from the slope).

## Oxygen p-doping of the buffer layer

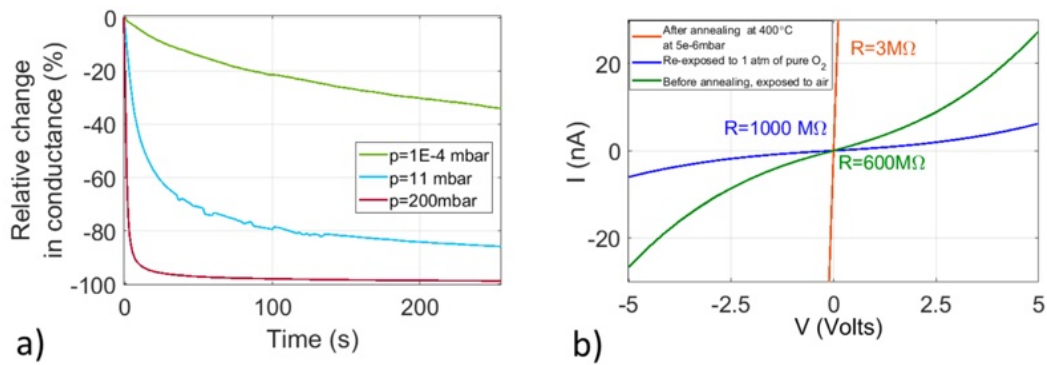

**Figure 4.11**: Effect of oxygen adsorption on a metal contacted buffer layer device (Fig. 4.1). a) Relative change of conductance after introduction of oxygen at time t=0s. b) IV curves of the devices in different conditions.

A survey is performed to identified some of the gases influencing the buffer layer devices uncovered by any protective layer. The devices are first annealed in vacuum to desorb the loosely bound contaminants. Upon cooling down to room temperature, a constant bias voltage is applied to the two-terminal device and the current is measured with a lock-in over time. A certain pressure of pure gas is then introduced into the chamber as the current is still measured over time. The introduction of 99.999% pure nitrogen, helium, or hydrogen at room temperature does not affect the conductance of the devices at pressures up to several hundreds of millibar. However, Fig. 4.11a shows the relative change in conductance over time of a metal contacted buffer layer device when a pressure of 99.999% pure oxygen gas is introduced in the transport measurement chamber (probe station). Qualitatively, the rate of the process is an increasing function of the oxygen pressure.

Fig 4.11b shows the IV curves of the same device in air, after annealing in vacuum, and after exposure to 1 atm of oxygen until saturation of the conductance. The inset indicates the resistance at zero bias for each curve. The resistance is the highest for the sample exposed to 1 atm of pure oxygen. This shows that oxygen is most likely the main component of air influencing the behavior of buffer layer devices exposed to ambient conditions.

### 4.3.2 Gated devices

The change in conductance of buffer layer devices in response to the adsorption of gases indicates that the Fermi level is not pinned and can shift. In this section, we address the field effect in the devices by fabricating and measuring top gated buffer layer devices. A thin film of $Al_2O_3$ is deposited on top of SWGNR contacted buffer layer devices, as described in section 4.1.3. The film thickness is about 30 nm in total (10 nm deposited by thermal evaporation and 20 nm of ALD). A trilayer Al/Pd/Au is deposited and patterned on top of the devices to produce the gate metal. IV curves are recorded for different gate voltages and temperature. Fig. 4.13 shows how the current changes with the applied gate voltage in a top gated SWGNR contacted buffer layer device for temperatures between 50 K and 300 K. The bias voltage is held constant at 1 V. The measurement is performed in DC conditions. For temperatures under 50 K, the conduction is below the minimum current measurable by the Keithley 2400. A positive voltage applied on the gate enhances the current while a negative voltage reduces it, which is consistent with electrons being the charge carriers in the device. The highest current ratio between $V_{gate}$ = -10 V and $V_{gate}$ = 10 V is $I_{on/off}$ = 20 at 100 K. The electron mobility µ can also be estimated using the relation:

$$\mu = \frac{\partial I}{\partial V_{gate}} \frac{1}{V_{bias}} \frac{L}{W} \frac{1}{C_i}$$

where L and W are the channel length and width, and $C_i$ is the capacitance per unit area of the gate dielectric. In our case, $C_i$ = 29.5 nFcm$^{-2}$ for 30 nm thick $Al_2O_3$. The device is 1.2 µm long and 5 µm wide. From the measurement at 300K in Fig. 4.13, **this gives a mobility µ = 0.97 cm$^2$V$^{-1}$s$^{-1}$**. As a comparison, the monolayer graphene grown on top of the buffer layer has a typical mobility µ > 1000 cm$^2$V$^{-1}$s$^{-1}$. This poor mobility combined with a relatively low on/off ratio makes the as grown buffer layer a less than ideal candidate for transistor applications. The origin of the low mobility of the electrons is investigated in the rest of this chapter. [The transport is found to be dominated by variable range hopping, consistent with considerable disorder].

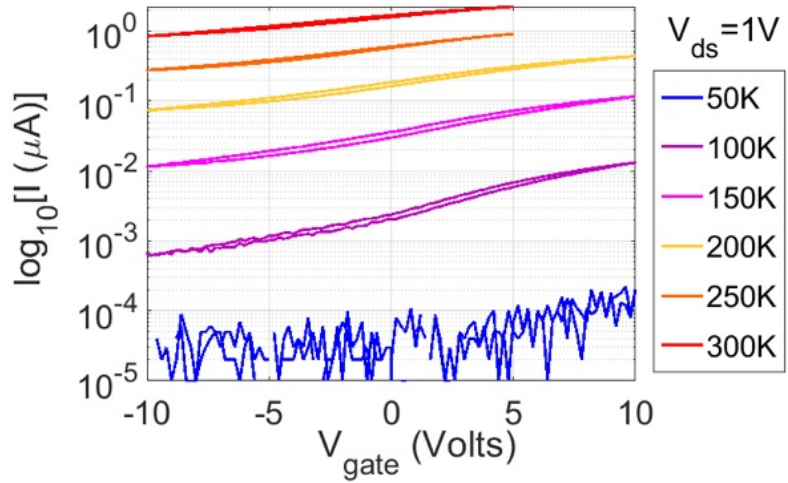

**Figure 4.13**: Log plot of the current versus gate voltage for $V_{bias}$ = 1 V, measured at various temperatures.

3. Palmer, J., et al., *Controlled epitaxial graphene growth within removable amorphous carbon corrals.* Applied Physics Letters, 2014. **105**(2).
4. Goler, S., et al., *Revealing the atomic structure of the buffer layer between SiC(0001) and epitaxial graphene.* Carbon, 2013. **51**: p. 249-254.
5. Rutter, G.M., et al., *Imaging the interface of epitaxial graphene with silicon carbide via scanning tunneling microscopy.* Physical Review B, 2007. **76**(23): p. 235416.
6. Prudkovskiy, V.S., et al., *An epitaxial graphene platform for zero-energy edge state nanoelectronics.* Nature Communications, 2022. **13**(1): p. 7814.
15